\newcommand{\forecai}[1]{#1}
\newcommand{\forarxiv}[1]{}

\forecai{%
\documentclass{ecai}
\ecaisubmission
}

\forarxiv{%
\documentclass{article}
}

\usepackage{times}  
\usepackage{helvet} 
\usepackage{courier}  
\usepackage[hyphens]{url} \usepackage{enumitem}
\usepackage{amssymb}  

\usepackage{graphicx} 
\usepackage{graphicx}  
\forecai{%
\setcounter{secnumdepth}{0}
}
\usepackage{amsmath,amsfonts}

\makeatletter
\renewcommand*{\@opargbegintheorem}[3]{\trivlist
      \item[\hskip \labelsep{\bfseries #1\ #2}] \textbf{(#3)}\ \itshape}
\makeatother
\usepackage{nicefrac}
\usepackage{xcolor}
\usepackage[english]{babel}
\usepackage{graphicx}
\usepackage{url}

\forarxiv{%
\newtheorem{theorem}{Theorem}
}
\newtheorem{corollary}{Corollary}

\newtheorem{observation}{Observation}
\newtheorem{lemma}{Lemma}

\newtheorem{definition}{Definition}

\newtheorem{example}{Example}
\newtheorem{remark}{Remark}

\newenvironment{proof}{\noindent\textbf{Proof.\quad}}{\hfill$\square$\newline}

\newcommand{\mypara}[1]{\smallskip\noindent\textbf{#1.}}

\newcommand{\qqed}{\hfill$\square$}

\newcommand{\calA}{\mathcal{A}}
\newcommand{\calC}{\mathcal{C}}
\newcommand{\calH}{\mathcal{H}}

\newcommand{\iti}{{\it i}}
\newcommand{\itii}{{\it ii}}
\newcommand{\itiii}{{\it iii}}

\newcommand{\HC}{{H}}

\begin{document}

\pagestyle{plain}

\title{Genuine Personal Identifiers and Mutual Sureties for Sybil-Resilient Community Formation}

\author{
Gal Shahaf$^1$,
Ehud Shapiro$^1$ \&
Nimrod Talmon$^2$\\
%\affiliations
$^1$Weizmann Institute of Science\\
$^2$Ben-Gurion University\\
%\emails
\{gal.shahaf, ehud.shapiro\}@weizmann.ac.il,
talmonn@bgu.ac.il
}

\pagestyle{plain}

\maketitle

\begin{abstract}
%
%Providing genuine identity to all people by 2030 is a UN Sustainable Development Goal. Yet, current top-down digital identity-granting solutions are unlikely to close the 1Bn-people gap in time, as they are not working for citizens of failed states  nor for people fleeing harshness.   While more than half the world population is now online, the prevailing types of digital identities provided online may be fake or duplicated, resulting in lack of accountability and trust. 
%
While most of humanity is suddenly on the net, the value of this singularity is hampered by the lack of credible digital identities:   Social networking,  person-to-person transactions, democratic conduct, cooperation and philanthropy are all hampered by the profound presence of fake identities, as illustrated by Facebook's removal of 5.4Bn fake accounts since the beginning of 2019. 

Here, we introduce the fundamental notion of a \emph{genuine personal identifier}---a globally unique and singular identifier of a person---and present a foundation for a decentralized, grassroots, bottom-up process in which every human being may create, own, and protect the privacy of a genuine personal identifier. 
The solution employs mutual sureties among owners of personal identifiers,  resulting in a mutual-surety graph reminiscent of a web-of-trust. Importantly, this approach is designed for a distributed realization, possibly using distributed ledger technology, and does not depend on the use or storage of biometric properties. For the solution to be complete, additional components are needed, notably a mechanism that encourages honest behavior~\cite{seuken2014sybil} and a sybil-resilient governance system~\cite{SRSC}.
\end{abstract}

\maketitle

\section{Introduction}\label{sec: intro}

Providing credible identities to all by 2030 is a UN Sustainable Development Goal. Yet, current top-down digital identity-granting solutions are unlikely to close the 1Bn-people gap~\cite{worldbankundocumented} in time, as they are not working for citizens of failed states  nor for people fleeing physical or political harshness~\cite{geisler2017impediments,humantide}.   Concurrently, humanity is going online at an astonishing rate, with more than half the world population now being connected. Still, online accounts do not provide a solution for credible digital identity either, as they may easily be fake, resulting in lack of accountability and trust. For example, Facebook reports the removal of 5.4Bn (!) fake accounts since the beginning of 2019~\cite{fung_2019,mccarthy_2019}.

The profound penetration of fake accounts on the net greatly hampers its utility for credible human discourse and any ensuing  deliberations and democratic decision making; it makes the net unsuitable for vulnerable populations, including children and the elderly; it makes the use of the net for person-to-person transactions, notably direct philanthropy, precarious; and in general it turns the net into an inhuman, even dangerous, ecosystem. As an aside, we note that the panacea of cryptocurrencies for the lack of credible personal identities on the net is the reckless employment of the environmentally-harmful proof-of-work protocol. 
\forecai{%
\newpage
}
\textbf{Our aim is a conceptual and mathematical foundation for allowing every person to create, own, and protect the privacy of a globally-unique and singular identifier}, henceforth referred to as \emph{genuine personal identifier}. We believe that successful deployment and broad adoption of genuine personal identifiers will afford solutions to these problems and more, providing for: egalitarian digital democratic governance in local communities and in global movements; digital cooperatives and digital credit unions; direct philanthropy; child-safe digital communities; preventing unwanted digital solicitation; banishing deep-fake (by marking as spam videos not signed by a genuine personal identifier);  credible and durable digital identities for people fleeing political or economic harshness; accountability for criminal activities on the net; and egalitarian cryptocurrencies employing an environmentally-friendly Byzantine-agreement consensus protocol among owners of genuine personal identifiers. Furthermore, genuine personal identifiers may provide the necessary digital foundation for a notion of \emph{global citizenship} and, subsequently, for democratic global governance~\cite{shapiro2018point}. 

Granting an identity document by a state is a complex process as it requires careful verification of the person's credentials. The process culminates in granting the applicant a state-wide identifier that is unique (no two people have the same identifier) and singular (no person has two identifiers).
Granting a genuine personal identifier might seem even more daunting, as it needs to be globally-unique, not only state-wide unique, except for following fundamental premise:\\ \emph{\textbf{Every} person deserves a genuine personal identifier}. Thus, there are no specific credentials to be checked, except for the existence of the person. As a result, a solution for all people to create and own genuine personal identifiers may focus solely on ensuring the one-to-one correspondence between people and their personal identifiers.

A solution that is workable for all must be decentralized, distributed, grassroots, and bottom-up. Solid foundations are being laid out by the notion of self-sovereign identities~\cite{ssi} and the W3C Decentralized Identifiers~\cite{did} and Verifiable Claims~\cite{vc}
emerging standards, which aim to let people freely create and own identifiers and associated credentials.  We augment this freedom with the goal that each person declares exactly one identifier as her \emph{genuine personal identifier}.  We note that besides the genuine personal identifier, one may create, own and use any number of identifiers of other types.

Becoming the owner of a genuine personal identifier is simple.  With a suitable app, this could be done with a click of a button:
\begin{enumerate}
    \item Choose a new cryptographic key-pair $(v,v^{-1})$, with $v$ being the public key and $v^{-1}$ the private key. % Keep $v^{-1}$ secure and secret!
    \item Claim $v$ to be your genuine personal identifier by publicly posting a declaration that $v$ is a genuine personal identifier, signed with $v^{-1}$. 
\end{enumerate}
Lo and behold! You have become the proud rightful owner of a genuine personal identifier.\footnote{As the public key may be quite long, one may also associate oneself with a shorter ``nickname'', a hash of the public key, e.g. a 128-bit hash (as a UUID) or a 256-bit hash (as common in the crypto world).}
Note that a declaration of a genuine personal identifier, by itself, does not reveal the person making the declaration; it only reveals to all that someone who knows the secret key for the public key $v$ claims $v$ as her genuine personal identifier. Depending on personality and habit, the person may or may not publicly associate oneself with~$v$.
E.g., a person with truthful social media accounts may wish to associate these accounts with its newly-minted genuine personal identifier.

\begin{figure*}[t]
\centering
\includegraphics[width=15cm]{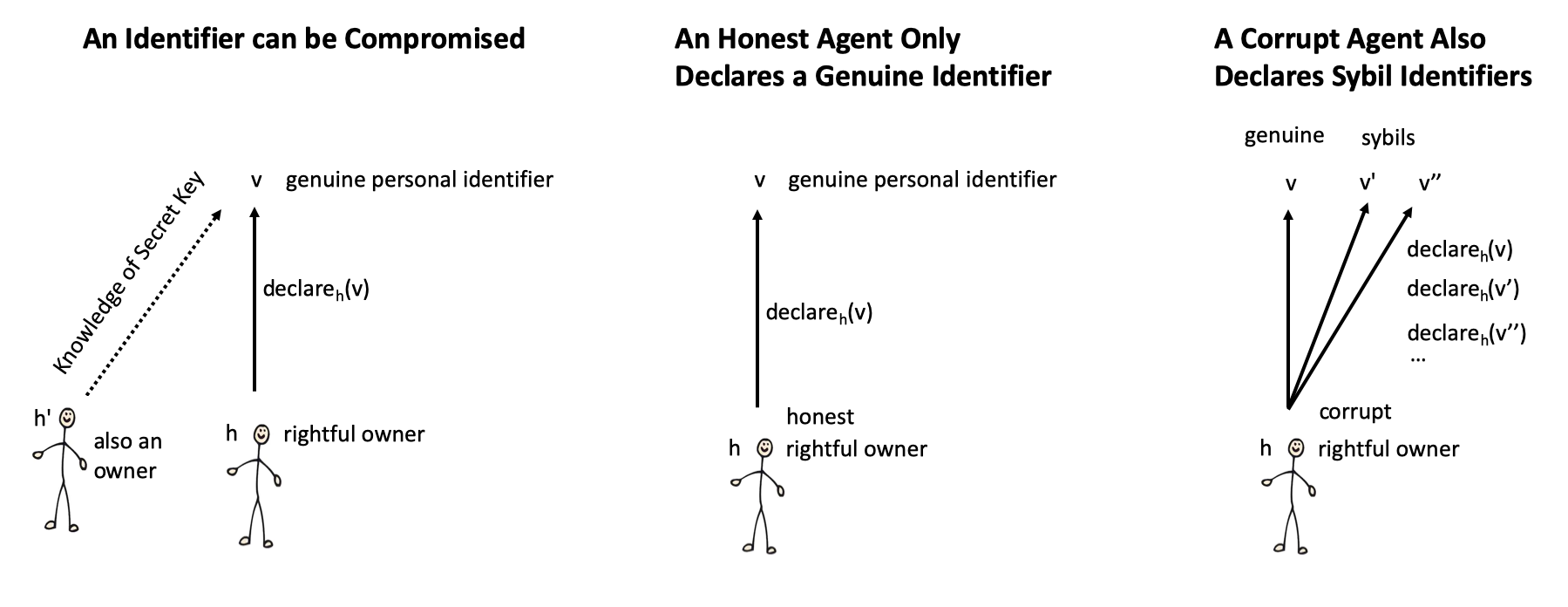}
\caption{A compromised identifier (left), honest agent (middle), and corrupt agent (right).} \label{figure:figureone}
\end{figure*}

If becoming the rightful owner of a genuine personal identifier is so simple, what could go wrong? In fact, so many things can go wrong, that this paper is but an initial investigation into describing, analyzing, and preventing them. Some of them are enumerated below; $h$ denotes an agent:
\begin{enumerate}
    
    \item  The key-pair $(v,v^{-1})$ is not new, or else someone got hold of it between Step 1 and Step 2 above. Either way, someone else has declared $v$ to be a genuine personal identifier prior to the declaration by $h$.  In which case $h$ cannot declare $v$. 
    
    \item Agent $h$ failed to keep $v^{-1}$ secret so that other people, e.g. $h'$, know $v^{-1}$, in which case $h'$ is also an \emph{owner} of~$v$ and, thus, $v$ is \emph{compromised}.
     Figure~\ref{figure:figureone} (left) illustrates a compromised personal identifier.
     
    \item %The private key $v^{-1}$ has been lost, stolen, robbed, or otherwise compromised. 
    The agent $h$ intended to divulge his association with his public key $v$ only on a need-to-know basis, but the association of $v$ and $h$ has become public knowledge, prompting agent $h$ to replace his genuine personal identifier $v$ with a new one.

    \item Agent $h$ declared $v$ as his genuine personal identifier, but later also declared another personal identifier $v'$. Then, $v$ and $v'$ are \emph{duplicates}, $v'$ is a \emph{sybil}, and agent $h$ is \emph{corrupt}.  An \emph{honest} agent does not declare sybils.
    Figure~\ref{figure:figureone} illustrates honest and corrupt agents.

\end{enumerate}

We aim to develop the foundation for genuine personal identifiers utilizing basic concepts of public-key cryptography, graph theory,  social choice theory and game theory. Here, we focus on utilizing the first two disciplines; see \cite{SRSC} for sybil-resilient social choice, which provides a foundation for democratic governance of digital communities with bounded sybil penetration; incorporating sybils in cooperative game theory, with the goal of forming a sybil-free grand coalition, is work in progress.

\mypara{Related Work}
Sybil-resilience has received considerable attention in AI research \cite{conitzertwo,andersen2008trust,borge2017proof,Borge2017ProofofPersonhoodRP,conitzer2010false,conitzer2010using,conitzerone,SRSC,waggoner2012evaluating,wagman2008optimal} as elaborated below.

Philosophically, genuine personal identifiers aim to bridge the gap between agents and their corresponding identifiers. This distinction, acknowledged in semiotics as the difference between \textit{signified} and \textit{signifier}, strongly relates to the study of sense and reference initiated by Frege \cite{frege2003sense} followed by vast literature in analytic philosophy and the philosophy of language. Conceptually, the formal framework suggested here may be viewed as an attempt to computationally realize unique and singular signifiers of agents in a distributed setting. 

Practically, digital identities are a subject of extensive study, with many organizations aiming at providing solutions, notably Self-Sovereign Identifiers~\cite{ssi}.
Business initiatives include the Decentralized Identity Foundation (\url{identity.foundation}),  the Global Identity Foundation~\cite{gif} and Sovrin~\cite{sovrin}. None of these projects are concerned with the uniqueness or singularity of personal identities.

Other high-profile projects to provide nationwide digital identities include India's Aadhaar system~\cite{aadhaar}, Sierra Leone's Kiva identity protocol \cite{staats2013kiva} and the World Food Programme’s %blockchain-based 
cash aid distribution program in refugee camps \cite{kshetri2018blockchain}. Here we are concerned with self-sovereign and global personal identities, not bound to any national boundaries or entities, and argue that top-down approaches fail to provide such a solution.
In this context, we mention the concept of Proof of Personhood~\cite{borge2017proof}, aiming at providing unique and singular identities by means of conducting face-to-face encounters, an approach suitable only for small communities. 

Our solution is based upon the notion of trust, thus we mention Andersen et al.~\cite{andersen2008trust}, studying axiomatizations of trust systems. They are not concerned, however, with sybils, but with quality of recommendations. 
We mention work on sybil-resilient community growth~\cite{poupko2019sybil}, describing algorithms for the growth of an online community that keep the fraction of sybils in it small; and work on sybil-resilient social choice~\cite{SRSC}, describing aggregation methods to be applied in situations where sybils have infiltrated the electorate.
In these two papers, a notion of genuine identities and sybils  is used without specifying what they are; here, we define a concrete notion of genuine personal identifiers, and derive from it a formal definition of sybils and related notions of honest and corrupt agents and byzantine identifiers.
Finally, we mention the work of Conitzer ~\cite{conitzerone,conitzertwo} regarding computerized tests for sybil fighting:
  A test that is hard for a person to pass more than once~\cite{conitzerone} and a test that is hard to pass simultaneously by one person~\cite{conitzertwo}.

\section{Genuine Personal Identifiers}
\mypara{Ingredients}
The ingredients needed for a realization of genuine personal identifiers are:

\begin{enumerate}
\item A set of agents.  It is important to note that, mathematically, the agents form a set (of unique entities) not a multiset (with duplicates). Intuitively, it is best to think of agents as people (or other physical beings with unique personal characteristics, unique personal history, and agency, such as intelligent aliens), which cannot be duplicated, but not as software agents, which can be.
  \item A way for agents to create cryptographic key-pairs. This can be realized, e.g., using the RSA standard~\cite{rsapaper}.
  Our solution does not require a global standard or a uniform implementation for public key encryption:
    Different agents can use different technologies for creating and using such key-pairs, as long as the signatures-verification methods are declared.
  
  \item A way for agents to sign strings using their key-pairs.  As we assume cryptographic hardness,  an agent that does not know a certain key-pair cannot sign strings with this key-pair.
  
  \item A bulletin board or public ledger, to which agents may post and observe signed messages, where all agents observe the same order of messages.  The weaker requirement that the same order is  observed only eventually, as is standard with distributed ledger protocols, could also be accommodated.  Considering partial orders is a subject of future work.

\end{enumerate}

\mypara{Agents and their Personal Identifiers}\label{subsection:people}
 We assume a set of agents $\calH$ that is fixed over time.\footnote{Birth and death of agents will be addressed in future work.} Agents can create new key-pairs $(v,v^{-1})$. We assume that an agent that has a key-pair can sign a string, and denote by $v(x)$ the string resulting from signing the string $x$ with $v^{-1}$.
Intuitively, each agent corresponds to a human being. Importantly, members of the set $\calH$ of agents (e.g., containing all human beings) cannot be referenced explicitly and, in particular, posted signed messages never refer directly to agents $h \in \calH$.
A key motivation for our work is providing people with digital genuine personal identifiers without accessing any of their intrinsic (e.g. biometric) private properties and without depending upon such properties as identifiers.

As we aim personal identifiers to be self-sovereign identities that conform to the W3C Decentralized Identifiers emerging standards, we let agents create and own their personal identifiers. An agent $h$ can publicly declare a personal identifier $v$ for which it knows the private key $v^{-1}$.  A \emph{personal identifier declaration} has the form $v(gid(v))$ and can be effected by agent $h$ posting $v(gid(v))$ to a public ledger. We denote this action by $declare_h(gid(v)) \in \calC$.  Recall that all agents have the same view of the sequence of all declarations made;  subsequent work may relax this assumption. 

\begin{definition}[Personal Identifier]\label{definition:global-identifier}
Let $\calC$ be a sequence of personal identifier declaration events and $declare_h(gid(v)) \in \calC$ the first declaration event in which $v$ occurs. Then $v$ is a \emph{personal identifier} and $h$ is the  \emph{rightful owner} of $v$, given $\calC$. 
\end{definition}

\begin{definition}[Genuine Personal Identifier, Sybil, Honest, and Corrupt Agents]\label{definition:genuine-global-identifier}
Let $\calC$ be a sequence of personal identifier declaration events and $h$ be the rightful owner of personal identifier $v$ in $\calC$.
Then $v$ is \emph{genuine} if it is the first personal identifier declared in $\calC$ by $h$, else $v$ is a \emph{sybil}.  An agent $h$ is \emph{corrupt} if it declares any sybils, else $h$ is \emph{honest}. (All notions are relative to~$\calC$.)
\end{definition}

See Figure~\ref{figure:figureone} and some remarks:
(1) An agent is the rightful owner of its genuine personal identifier as well as of any subsequent sybils that it declares.
(2) If $h$, the rightful owner of $v$, is corrupt, then its first declared identifier is genuine and the rest of its declared identifiers are all sybils.
(3) An honest agent may create and use many key-pairs for various purposes, yet remain honest as long as it has declared at most one public key as a personal identifier.

\subsection{Mutual Sureties and Their Graphs}

A key element of our approach is the pledging of mutual sureties by agents.
Intuitively, mutual surety pledges provide a notion of trust between the owners of personal identifiers.  They allow two agents that know each other and know the personal identifiers declared by each other to vouch for the other regarding the good standing of the personal identifiers.  This notion is key to help honest agents fend-off sybils. We consider several types of mutual sureties, of increasing strength, and illustrate the corresponding sybil-resilience each type of mutual sureties can obtain.

For an agent, say $h$, to provide surety regarding the personal identifier of another agent, say $h'$, $h$ first has to know $h'$.  How this knowledge is established is not specified in our formal framework, but this is quite an onerous requirement that cannot be taken lightly or satisfied casually. 
E.g., we may assume that one knows one's family, friends and colleagues, and may diligently get to know new people if one so chooses.
%
% Both parties to a violated surety should be punished, but not necessarily in the same way.  For example, if a mutual surety with a sybil is discovered, the sybil should be removed and all agents pledging sureties with this sybil should be prosecuted. The punishment of the agent pledging for a sybil may include a fine and a temporary or permanent ban on pledging further sureties.  In addition,  all  existing sureties of the pledging agent should be examined for possible further violations.
%
%If $h'$ knows $h$, then $h$ can prove to $h'$ knowledge of $ kp(p,s)$, as discussed above, by $h$ signing with $s$ a challenge string provided by $h'$, which $h'$ can then verify using $p$.  To prevent $h$ from repudiating ownership of $p$ (e.g. when a duplicate identifier owned by $h$ is discovered), the disclosure of $p$ by $h$ can be videotaped and the challenge string signed by $s$ be kept.
%
We consider several types of sureties of increasing strength, in which an agent $h$ with personal identifier $v$ makes a pledge regarding the personal identifier $v'$ of another agent $h'$; all assume that the agent $h$ knows the agent $h'$.
We describe four \emph{Surety Types}, which are cumulative as each includes
all previous ones,  explain on what basis one may choose to pledge each of them, and present results that utilize them,
\begin{description}
 
    \item[Surety of Type 1:  \emph{Ownership} of a personal identifier.] Agent $h$ pledges that agent $h'$ owns personal identifier $v'$. 

\end{description}

Agent $h'$ can prove to agent $h$ that she owns $v'$ without disclosing $v'^{-1}$ to $h$.  This can be done, for example, by $h$ asking $h'$ to sign a novel string $x$ and verifying that $v'(x)$ is signed using $v'^{-1}$. 
This surety type is the weakest of all four, it is the one given in ``key signing parties'', and is implicitly assumed by applications such as PGP and web-of-trust \cite{abdul1997pgp}. For a given surety type, we say that the surety is \emph{violated} if its assertion does not hold; in particular, a surety of Type 1 is violated if $h'$ in fact does not know the secret key $v'^{-1}$. 

In general, mutual surety between two agents with two personal identifiers is pledged by each of the two agents pledging a surety to the personal identifier of the other agent.\footnote{We consider undirected graphs, as we require surety to be symmetric. Indeed, one may consider directed sureties.}  We define below three additional surety types, where the format of a surety pledge of Type $X$ by the owner of $v$ to the owner of $v'$ is $v(suretyX(v'))$, $X \in \{1,2,3,4\}$. The corresponding surety event is $pledge_{h}v(suretyX(v'))$, and the surety enters into effect once both parties have made the mutual pledges.  We now take $\calC$ to be a record  of both declaration events and pledge events.

\begin{definition}[Mutual Surety]\label{definition:mutual-surety}
The personal identifiers $v, v'$ have \emph{mutual surety of type X}, $X \in \{1,2,3,4\}$, if there are $h, h' \in \calH$ for which 
$pledge_{h}v(suretyX(v')) \in \calC \And 
pledge_{h'}v'(suretyX(v)) \in \calC,$
in which case $h$ and $h'$ are the \emph{witnesses} for the mutual surety between $v$ and $v'$.
\end{definition}

A sequence of events induces a sequence of surety graphs in which the vertices are personal identifiers that correspond to personal identifier declarations and the edges correspond to mutual surety pledges.
\begin{definition}[Surety Graph]\label{def: surety graph}
Let $\calC = c_1, c_2, \ldots$ be a sequence of events and let $\calC_k$ denote its first $k \ge 0$ events.
Then, for each $k\ge 0$,  $\calC_k$ induces a \emph{surety graph of type X}, $GX_k = (V_k, EX_k)$, $X \in \{1,2,3,4\}$, as follows:\footnote{We allow surety pledges to be made before the corresponding personal identifier declarations, as we do not see a reason to enforce order.}
$$ V_k = \{v ~|~ declare_h(gid(v)) \in \calC_k \textrm{ for some } h\in \calH\} $$
\begin{align*}
  EX_k = \{ (v,v') ~|~& pledge_h v(suretyX(v')) \in \calC_k,\\
  &pledge_{h'} v'(suretyX(v)) \in \calC_k,\\
  &\textrm{for some } h, h' \in \calH, v, v' \in V_k\} 
\end{align*}
\end{definition}

\begin{remark}
    Observe that mutual sureties can be easily pledged by agents, technically. However, we wish agents to be prudent and sincere in their mutual surety pledges. Thus, we expect a mechanism that, on one hand,  rewards the pledging of sureties but, on the other hand, punishes for surety violations, for example based on the approach of \cite{seuken2014sybil}. While the specifics of such a mechanism is beyond the scope of the current paper, note that with such a mechanism in place, the commissive illocutionary force~\cite{illocutionarybook} of a surety pledge will come to bear.
\end{remark}

\section{Updating a Personal Identifier with Mutual Sureties}\label{section:update-identifier}

  Once creating a genuine personal identifier is provided for, one must also consider the many circumstances under which a person may wish to update their personal identifier:
\begin{enumerate}
    \item \textbf{Identifier loss:} The private key was lost.
    \item \textbf{Identifier theft:} The private key was stolen, robbed, extorted, or otherwise compromised.
    \item \textbf{Identifier breach of privacy:}  The association between the personal identifier and the person was accidentally or maliciously disclosed with unwarranted consequences.
    \item \textbf{Identifier refresh:} Proactive identifier update to protect against all the above.
\end{enumerate}

\mypara{Update in Case of Loss or Theft}
The personal identifier declaration event $declare_h(gid(v))$ establishes $v$ as a personal identifier. To support updating a personal identifier, we add the personal identifier update event $declare_h(gid(v,v'))$, which declares that $v'$ is a new personal identifier that replaces $v$.  A public declaration of identifier update has the form $v(gid(v,v'))$, i.e., it is signed with the new identifier.
We refer to declarations of both types as \emph{personal identifier declarations}, and extend the assumption that a new identifier can be declared at most once to this broader definition of identifier declaration.
The \emph{validity} of an identifier update declaration is defined inductively, as follows.

\begin{definition}[Valid identifier Update declaration]
Let $\calC$ be a sequence of declarations, $V$ the set of global identities declared in $\calC$, and $h \in \calH$.  A \emph{personal identifier update event} over $V$ has the form $declare_h(gid(v,v'))$, $v, v' \in V$.

A personal identifier update event $declare_h(gid(v,v')) \in \calC$ is \emph{valid} and $h$ is the \emph{rightful owner} of $v$ if it is the first identifier declaration event of $v$ and $h$ is the rightful owner of $v'$.
\qqed \end{definition}
  Valid personal identifier declarations should form linear chains, one for each agent, each starting from $gid(v)$ and ending with the currently valid personal identifier of the agent: 
\begin{definition}[Identifier Provenance Chain]
Let  $\calC$ be a sequence of declarations and $V$ the declared set of global identities.
An \emph{identifier provenance chain} (\emph{provenance chain} for short) is a subsequence of $\calC$ of  the form (starting from the bottom):
\begin{align*}
&declare_{h_k}(gid(v_k,v_{k-1})),\\ &declare_{h_{k-1}}(gid(v_{k-1},v_{k-2})), \\ &\ldots \\
&declare_{h_1} (gid(v_1)).
\end{align*}
Such a provenance chain is \emph{valid} if the declarations in it are valid. Such a provenance chain is \emph{maximal} if there is no declaration $$declare_h(gid(v,v_k)) \in \calC$$ for any $v \in V$ and $h \in \calH$. A personal identifier $v$ is \emph{current} in $\calC$ if it is the last identifier $v = v_k$ in a maximal provenance chain in $\calC$.
\qqed \end{definition}

Note that it is very easy for an agent to make an  update declaration for its identifier.  However, it is just as easy for an adversarial agent wishing to steal the identifier to make such a declaration.  Hence, this ability must be coupled with a mechanism that protects the rightful owner of an identifier from identifier theft through invalid identifier update declarations.
Here we propose to use a stronger type of mutual sureties to support valid identifier update declarations and help distinguish between them and invalid declarations. 

\begin{description}
 
    \item[Surety of Type 2:]\textbf{\emph{Rightful} ownership of a personal identifier.} 
    Agent $h$ pledges that $h'$ is the rightful owner of personal identifier~$v'$. 

\end{description}
In addition to proving to $h$ that it owns $v'$, $h'$ must provide evidence that $h'$ itself, and not some other agent, has declared $v'$.  A selfie video of $h'$ pressing the \emph{declare} button with $v'$, signed with a certified timestamp promptly after the video was taken, and then signed by $v'$, may constitute such evidence.  A suitable app may record, timestamp, and sign such a selfie video automatically during the creation of a genuine personal identifier. In particular, this surety is violated if $h'$ in fact did not declared $v'$ as a personal identifier. 

Note that immediately following an identifier update declaration, the new identifier may not have any surety edges incident to it. Thus, as a crude measure, we may require that the identifier update would come to bear only after all the Type 2 surety neighbors of the old identifier, or a sufficiently large majority of them,  would update their mutual sureties to be with the new identifier.  To achieve that, an agent wishing to update its identifier would have to approach its neighbors and to create such updated Type 2 mutual surety pledges. 

\begin{example}
Consider two friends, agent $h$ and agent $h'$ having a mutual surety pledge between them. If $h'$ would lose her identifier, she would create a new key-pair, make an identifier update declaration, and ask $h$ for a new mutual surety pledge between $h$'s identifier and $h'$'s new identifier. 
\end{example}

The following observation follows from: (1) a valid provenance chain has a single owner; and (2) whether a Type 2 surety between two identifiers is violated depends on their rightful owners.

\begin{observation}\label{observation:import}
Let $\calC$ be a sequence of update declarations and  $C_1, C_2$ be two valid provenance chains in $\calC$.  If a Type 2 surety pledge between two global identities $v_1 \in C_1, v_2 \in C_2$ is valid, then any Type 2 surety pledge between two personal identifiers in these provenance chains, $u_1 \in C_1, u_2 \in C_2$ is valid.
\end{observation}

The import of Observation~\ref{observation:import} is that a Type 2 mutual surety can be ``moved along'' valid provenance chains as they grow, without being violated, as it should be.
Below we argue that invalid identifier update declarations are quite easy to catch, thus the risk of stealing identities can be managed. In effect, we show the value of Type 2 surety pledges in defending an identifier against theft via invalid update declarations.

Let $\calC$ be a sequence of declarations, $C_1, C_2$ be two provenance chains in $\calC$, and assume that there is a valid Type 2 surety pledge between the two current global identifiers $v_1 \in C_1, v_2 \in C_2$, made by $h_1 =$ \emph{Marry} and $h_2 =$ \emph{John}.  
Now assume that the identifier update declaration $c= declare_{h} (gid(v,v_1))$ is made, namely, some agent \emph{Sue} $\ne$ \emph{Marry} has declared to replace $v_1$ by $v$.  
Then, it will be hard for \emph{Marry} to secure surety from \emph{John} and, if she attempts to do so, then \emph{John} will know that $c$ is not valid and thus (if \emph{John} is honest) a Type 2 mutual surety between $v$ and $v_2$ will not be established. 
Consider the following case analysis:
\begin{itemize}
    \item 
    Assume \emph{Marry} notices $c$.  Then she would inform \emph{John} that she did not declare $c$, and thus \emph{John} will know that $c$ is not valid.
    
    \item
    Assume that \emph{John} notices $c$.  He would approach \emph{Marry} to update the Type 2 mutual surety between them accordingly; \emph{Marry} would deny owning $v$, and thus \emph{John} will know that $c$ is invalid.  
    
    \item
    Alternatively,  \emph{Sue} would approach \emph{John} to update the Type 2 mutual surety of $v_2$ with $v_1$ to be with $v$ instead; \emph{John} will see (or suspect, if \emph{Sue} did not reveal herself) that \emph{Sue} is not \emph{Marry}, will double check with \emph{Marry} and deem the declaration $c$ invalid.

\end{itemize}

\mypara{Reset in Case of Breach of Privacy}
Note that provenance chains address identifier update in case of loss of theft, but do not address breach of privacy, as the updated new identifier is publicly tied to the previous one.  To address breach of privacy, a person first has to invalidate his existing identifier, then create a new genuine identifier that is not linked to the previous one; this may result in loss of any public credit and goodwill associated with the old genuine personal identifier, but there may be circumstances in which a person would need to protect his privacy even at the expense of such loss.  To facilitate that, an identifier reset declaration uses the special value $\textbf{null}$ and has the form $declare_h(gid(v,\textbf{null}))$, which nullifies $v$ as a personal identifier, and enters into effect if supported by those who initially provided the surety to $v$. Agent $h$ would then be free to declare a new personal identifier $v'$, which is not tied to the now-defunct $v$, and at the same time without $v'$ being a sybil and without $h$ becoming corrupt as a result of this declaration.  

We note that using $\textbf{null}$, an initial genuine identifier declaration could be of the form 
$declare_h(gid(\textbf{null},v))$, thus dispensing with the unary declaration format $declare_h(gid(v))$ altogether.

\mypara{GDPR}
In general, our approach does not imply any ``data controllers'' or ``data processors''~\cite{GDPR} other than people and their trusted friends, but if realized on a large scale it might require such, possibly democratically-appointed by the large community that needs them. Furthermore, our approach and does not record or store any ``personally identifiable information'', and as such we believe is compliant with GDPR~\cite{GDPR}.
One exception is perhaps the association of a genuine personal identifier with the person that owns it, which could be made public on purpose by the owner, inadvertently  by the owner or by another person, or maliciously by another person. 
Closely tied to this exception is GDPR's ``right to be forgotten''~\cite[Article 17]{GDPR}, which is notoriously difficult to realize in a distributed setting and hence resulted in sweeping legal opinions regarding the inapplicability of distributed ledger/blockchain technology for the storage of personally identifiable information~\cite{BlockchainAndGDPR}.

Our method of personal identifier reset first invalidates an existing  identifier (which could be coupled with a demand of erasure of every reference to this identifier from any public data controller, e.g. Facebook) and then creates an independent and unlinked new personal identifier.  This may be the first proposal on how to realize the ``right to be forgotten'' in a decentralized setting.

\section{Sybil- and Byzantine-Resilient Community Growth}% with Mutual Sureties}

Ideally, we would like to attain sybil-free communities, but acknowledge that one cannot prevent sybils from being declared and, furthermore, perfect detection and eradication of sybils is out of reach. %
Thus, our aim is to provide the foundation for a digital community of genuine personal identifiers to grow by admitting new identifiers indefinitely, while retaining a bounded sybil penetration.
As noted above, democratic governance can be achieved even with bounded sybils penetration~\cite{SRSC}. %Sybil-safety, namely the inability of sybils to change the status quo against the will of the genuine agents, can be achieved with any bounded sybil-penetration.  Sybil-liveness, namely the inability of the sybils to block a change of the status quo by the genuine agents, can be achieved provided sybil penetration is below one third. 

\mypara{Community history} For simplicity, we assume a single global community $A$ and consider elementary transitions obtained by either adding a single member to the community or removing a single community member: 

\begin{definition}[Elementary Community Transition]
Let $A,A'$ denote two communities in $V$. We say that $A'$ is obtained from $A$
by an \emph{elementary community transition}, and we denote it by $A\rightarrow A'$, if:
\begin{itemize}
    \item $A' = A$, or
    \item $A' = A \cup \{v\}$ for some $v\in V\setminus A$, or
    \item $A' = A \setminus \{v\}$ for some $v\in A$.
\end{itemize}
\end{definition}

\begin{definition}[Community History]
    Let $\calC = c_1,c_2,...$ be a sequence of events. A \emph{community history} wrt.\ $\calC$ is a sequence of communities $\calA = A_1,A_2,\ldots$ such that $A_i\subseteq V_i$ and $A_i \rightarrow A_{i+1}$ holds for every $i\geq 1$.
\end{definition}

We do not consider community governance in this paper, only the effects of community decisions to add or remove members.  Hence, we assume that the sequence of events $\calC$ includes the events $add_A(v)$ and $remove_A(v)$.  With this addition, $\calC= c_1, c_2,\ldots$  induces a community history $A_1,A_2,...$, where $A_{i+1} = A_i\cup\{v\}$ if $c_i = add_A(v)$ for $v \in V \setminus A_i$; $A_{i+1} = A_i\setminus\{v\}$ if $c_i = remove_A(v)$ for $v\in A_i$; else $A_{i+1}$ = $A_i$.

\begin{definition}[Community, Sybil penetration rate]
Let $\calC$ be a sequence of events and 
let $S\subseteq V$ denote the sybils in $V$ wrt. $\calC$.
A \textit{community} in $V$ is a subset of identifiers $A\subseteq V$.  The \emph{sybil penetration} $\sigma(A)$ of the community $A$ is given by $\sigma(A) = \frac{|A\cap S|}{|A|}$.
\end{definition}

The following observation is immediate.
\begin{observation}\label{ob: naive}
Let $A_1,A_2,...$ be the community history wrt. a sequence of declarations $\calC$. Assume that $A_1\subseteq H$, and that whenever $A_{i+1} = A_i\cup\{v\}$ for some $v\notin A_i$, it holds that $Pr(v\in S)\leq \sigma$ for some fixed $0\leq \sigma \leq 1$. Then, the expected sybil penetration rate for every $A_i$ is at most $\sigma$.
\end{observation}
That is, the observation above states that a sybil-free community can keep its sybil penetration rate below $\sigma$, as long as the probability of admitting a sybil to it is at most $\sigma$.
While the simplicity of Observation \ref{ob: naive} might seem promising, its premise is naively optimistic. Due to the ease in which sybils can be created and to the benefits of owning sybils in a democratic community, the realistic scenario is of a hoard of sybils and a modest number of genuine personal identifiers hoping to join the community. Furthermore, once a fraction of sybils has already been admitted, it is reasonable to assume that all of them (together with their perpetrators of course) would support the admission of further sybils. Thus, there is no reason to assume neither the independence of candidates being sybils, nor a constant upper bound on the probability of sybil admission to the community.  
Hence, in the following we explore sybil-resilient community growth under more realistic assumptions.

\subsection{Sybil-Resilient Community Growth}\label{sec: sybils}

A far more conservative assumption includes a process employed by the community with the aim of detecting sybils. We shall use the abstract notion of \emph{sybil detector} in order to capture such process, that may take the form of a query, a data-based comparison to other identifiers, or a personal investigation by some other agent. To leverage this detector to sybil-resilient community growth regardless of the sybil distribution among the candidates, we shall utilize a stronger surety type, defined as follows:

\begin{description}
 
    \item[Surety of Type 3:] \textbf{Rightful ownership of a \emph{genuine} personal identifier.}
    Agent $h$ pledges Surety Type 2 and that $v'$ is the genuine personal identifier of $h'$.
\end{description}

Providing this surety requires a leap of faith.  In addition to $h$ obtaining from $h'$ a proof of rightful ownership of $v'$, $h$ must also trust $h'$ not to have declared any other personal identifier prior to declaring $v'$.  There is no reasonable way for $h'$ to prove this to $h$, hence the leap of faith. 

Since Type 3 sureties inherently aim to distinguish between genuine identifiers and sybils, sybil-resilient community growth is established upon the underlying $G3$ surety graph. Specifically, we consider a setting where potential candidates to join the community are identifiers with a surety obtained from current community members. Conversely, we consider a violation of a surety in one direction as a strong indication that the surety in the other direction is violated. That is, if $(v,v')\in E$ and $v'$ was shown to be sybil, (i.e., $h'$ has declared some other $v''$ as a personal identifier before declaring $v'$ as a personal identifier), then $v$ should undergo a thorough investigation in order to determine whether it is sybil as well. 

Next, we formalize this intuition in a simple stochastic model where admissions of new  members are interleaved with random sybil detection among community members:

\begin{enumerate}[label=(\arabic*)]
    
    \item An identifier is admitted to the community via an elementary community transition $A\rightarrow A\cup \{v\}$ only if there is some $a\in A$ with $(a,v)\in E$.
    
    \item Every admittance of a candidate is followed by a random sybil detection within the community: An identifier $a\in A$ is chosen uniformly at random. If $a$ is genuine it is declared as such. If $a$ is sybil, it is successfully detected with probability $0 < p \leq 1$. 
    
    \item The detection of a sybil implies the successful detection of its entire connected sybil component (with probability 1). That is, if $a$ is detected as sybil, then the entire connected component of $a$ in the sybil subgraph  $G|_S$ is detected and expelled from the community.
    
    \item The sybils are operated from at most $j \le k$ disjoint sybil components in $A$. Furthermore, we assume that sybils join sybil components uniformly at random, i.e., a new sybil member has a surety to a given sybil component with probability $\frac{1}{k}$, else, it forms a new sybil component with probability $\frac{k-j}{k}$.
    
\end{enumerate}

Note that assumption~(2) is far weaker than the premise in Observation~\ref{ob: naive} as it presumes nothing on the sybil penetration among the candidates, but rather on the proactive ability to detect a sybil, once examined. Assumption (3) exploits the natural cooperation among sybils, especially if owned by the same agent, and assumes that if a sybil is detected by a shallow random check with probability $p$, then all its neighbours will be thoroughly investigated and will be detected if sybil with probability $1$, continuing the investigation iteratively until the entire connected sybil component of the initially-detected sybil is identified. In assumption (4), the parameter $k$ and the locations of the components are adversarial -- the attacker may choose how to operate. While realistic attackers may also choose to which component shall the new (sybil) member join, uniformity is assumed to simplify the analysis. Possible relaxations of this model are future work. 

For this setting we show an upper bound on the expected sybil penetration, assuming bounded computational resources of the attacker.

\begin{theorem}\label{thm: steady state comp}
In the stochastic model described above, obtaining an expected sybil penetration $E[\sigma(A)] \geq \epsilon$ is NP-hard for every  $\epsilon>0$.
\end{theorem}

\begin{proof} Let $X_i\subseteq A$ denote a sybil component within the community at time $i$. In the stochastic model described above, $X_i$ is detected and immediately expelled with probability $p\cdot \frac{|X_i|}{|A_i|}$. The expected size of the component in this model is obtained in a steady state, i.e., in a state $i$ in time where $E[|X_{i+1}|] = |X_{i}|$, that is:
$$(1-px/n)\cdot\frac{1}{k}\cdot (x+1) + (1-px/n)\cdot(1-\frac{1}{k})\cdot x = x,$$
where $n:=|A_i|$ and $x=|X_i|$. It follows that $x^2 + \frac{1}{k} x -\frac{n}{pk} = 0$. Solving this quadratic equation implies that the size of a single sybil component in the steady state is $x \leq \sqrt{n/pk}$.
It follows that the number of sybils in the community in a steady state is $xk\leq \sqrt{nk/p}$.

The crucial observation now is that operating from $k$ nonempty sybil components corresponds to obtaining an independent set of size $k$ (at least, choosing a single vertex in each component). The theorem follows from the fact that approximating independent set within a constant factor is NP-hard (see, e.g.,~\cite{arora2009computational}).
\end{proof}

The following corollary establishes an upper bound on the sybil penetration rate regardless of the attacker's computational power. The result is formulated in terms of the second eigenvalue of the graph restricted to $A_i$. ($\lambda(G|_{A_i})$ is defined in the supplementary materials section.

\begin{corollary}\label{cor: steady state gen}
Let $\calA = A_1,A_2,\ldots$ be a community history wrt.\ a sequence of events $\calC$. If every community $A_i \in \calA$ with $A_i =  A_{i-1}\uplus\{v\}$ satisfies $\lambda(G|_{A_i})<\lambda$, then the expected sybil penetration in every $A_i\in \calA$ under the stochastic model depicted above, is at most $\sqrt{\lambda /p}$.
% and sybil penetration $\sigma(A_i)\le \beta - \gamma(1-\beta)$. 
\end{corollary}

\begin{proof}
Recall that the size of the maximal independent set is a trivial upper bound on $k$. The cardinality of an independent set in a $\lambda$-expander is at most $\lambda n$~\cite{hoory2006expander}; thus, $k\leq \lambda n$. It follows that the number of sybils in the community in a steady state is $xk\leq \sqrt{nk/p}\leq n\sqrt{\lambda /p}$.
\end{proof}

\subsection{Byzantine-Resilient Community Growth}

Here we consider the challenge of byzantine-resilient community growth. Intuitively, the term \emph{byzantines} aims to capture identifiers owned by agents that are acting maliciously, possibly in collaboration with other malicious agents. Formally, we define byzantines as follows.

\begin{definition}[Byzantine and harmless identifiers, Byzantine penetration]
An identifier is said to be \emph{byzantine} if it is either a sybil or the personal identifier of a corrupt agent. Non-byzantine identifiers are referred to as \emph{harmless}. We denote the byzantine and harmless identifiers in $V$ by $B,H \subseteq V$, respectively. 
The \emph{byzantine penetration} $\beta(A)$ of a community $A\subseteq V$ is given by $\beta(A) = \frac{|A\cap B|}{|A|}$.
\end{definition}
Since $S \subseteq B$, it holds that $\sigma(A) \le \beta(A)$ for every community $A$, hence an upper bound on the byzantine penetration also provides an upper bound on the sybil penetration.

As byzantine identifiers include genuine identifiers, they are inherently harder to detect, and thus the detection-based model described in Section \ref{sec: sybils} is no longer applicable in this setting. Rather, to achieve byzantine-resilient community growth, we rely on a stronger surety type, defined as follows:

\begin{description}
 
    \item[Surety of Type 4:] \textbf{Rightful ownership of a genuine personal identifier by an \emph{honest} agent.}
     Agent $h$ pledges Surety Type 3 and, furthermore, that $v'$ is a genuine personal identifier of an honest agent $h'$.
\end{description}

\begin{figure}[t]
\centering
\includegraphics[width=7cm]{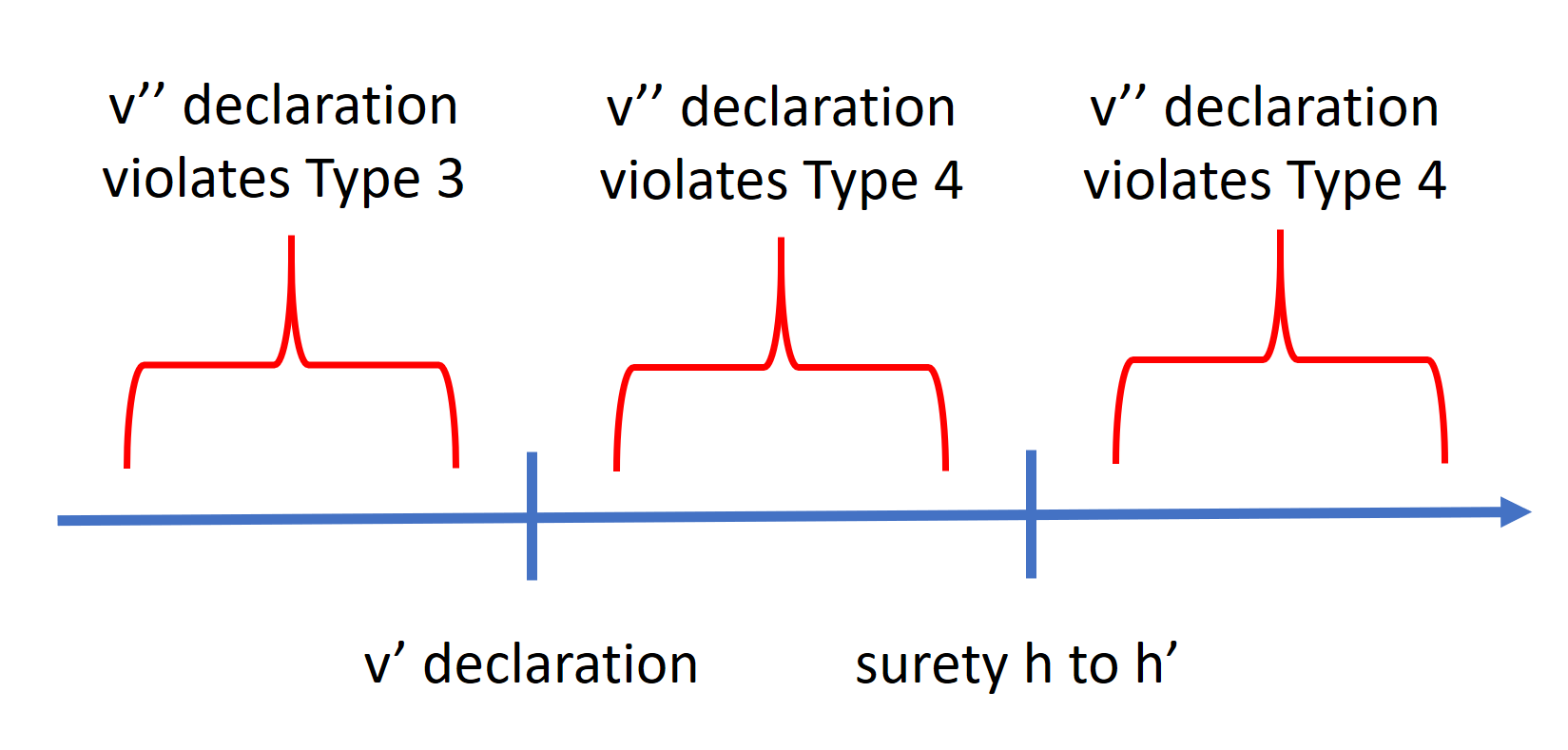}
\caption{Violations of Sureties of Type 3 and Type 4. The axis of time goes from left to right, with the points in time in which $v'$ was declared by $h'$ and the surety from $h$ to $h'$ was declared. Then, the braces describe the time regions in which, if another identifier $v''$ was to be declared by $h'$, would correspond to a violation of a surety of Type 3 or  4.} \label{figure:figuresurety}
\end{figure}

    Here $h$ has to put even greater trust in $h'$:  Not only does $h$ has to trust that the past actions of $h'$ resulted in $v'$ being her genuine personal identifier, but she also has to take on faith that $h'$ has not declared any sybils since and, furthermore, that $h'$ will not do so in the future. Note that a Type 4 surety is violated if after $h'$ declares $v'$ it ever declares some other $v''$ as a personal identifier. See Figure~\ref{figure:figuresurety} for illustrations of violations of sureties of Types 3 and 4.

% In our setting, these acts include declaring sybil identifiers and pledging violated sureties.

% Next we show how to obtain sybil-resilient community growth with sureties of Type 4 and a strengthened Type 3 surety. Two common concepts we use are byzantine identifiers and attack edges. Our key assumption is that the fraction of attack edges among all edges can be bounded; note that some assumption regarding the power of the attacker must be stated.  Together with a lower bound on the conductance of the surety graph and an upper bound on the fraction of surety edges devoted to identifiers outside the community, we can obtain a bound on the fraction of sybils in a growing community.

In the following, we provide sufficient conditions for Type 4 sureties to be used for byzantine-resilient community growth. We utilize the notation $A\twoheadrightarrow A'$ to indicate that $A'$ was obtained from $A$ via a finite sequence of elementary community transitions of incremental growth. Formally,  $A\twoheadrightarrow A'$ if there exists $k\in \mathbb{N}$ with $A = A_0 \rightarrow A_1\rightarrow...  A_k = A'$, with $|A_i \setminus A_{i-1}| = 1$ for all $i \in [k]$.

% Our key assumption is that attack edges are scarce, as honest people tend to trust honest people and distrust corrupt ones.

% \gal{ffff}

% Using the notion of $(\alpha,\beta,\gamma)$-community resilience defined by Poupko et al.~\cite{poupko2019sybil}, the following theorem guarantees byzantine-resilient community growth in this context.

\begin{theorem}\label{thm: byz frac}
Let $\calA = A_1 \twoheadrightarrow A_2 \twoheadrightarrow A_3 \twoheadrightarrow \ldots$ be a community history wrt.\ a sequence of events $\calC$. Set a sequence of degrees $d_1,d_2,\ldots$ and parameters $\alpha,\beta,\gamma,\delta \in [0,1]$. Assume:
\begin{enumerate}
    \item $deg(v)\le d_i$ for all $v\in A_i$, $i\in \mathbb{N}$.
    
    \item Every $a\in A_{i}$ satisfies $\frac{|\{x\in A_{i}~|~(a,x)\in E\}|}{d_i}\geq \alpha$.
    
    \item $\frac{|A_{1}\cap B|}{|A_{1}|}\le\beta$.
    
    \item $\frac{e(A_{i}\cap \HC,A_{i}\cap B)}{vol_{A_{i}}(A_{i}\cap \HC)}\leq \gamma$.
    
    \item $|A_{i}\setminus A_{i-1}| \leq \delta |A_{i-1}|$, with $\beta + \delta \leq \frac{1}{2}.$
    
    \item $\Phi(G|_{A_{i}}) > \frac{\gamma}{\alpha}\cdot \left(\frac{1-\beta}{\beta}\right)$.
    
\end{enumerate}

Then, every community $A_i\in \calA$ has Byzantine penetration $\beta(A_i) \le \beta$.
\end{theorem}

Roughly speaking, Theorem \ref{thm: byz frac} suggests that whenever: (1) Each graph $G|_{A_i}$ has a bounded degree $d_i$; (2) Sufficiently many edges are within $A_{i}$; (3) Byzantine penetration to $A_{1}$ is bounded; (4) Edges between harmless and byzantine identifiers are scarce; (5) Community growth in each step is bounded; (6) The conductance within $G|_{A_{i}}$ is sufficiently high; Then, the community may grow indefinitely with bounded byzantine penetration. 

As Theorem \ref{thm: byz frac} allows byzantine resilient growth at a fixed rate $\delta$, it may be interpreted as an extension of a related result by Poupko et al.~\cite{poupko2019sybil}, where new members were added one at a time. Moreover, ~\cite{poupko2019sybil} assumed a notion of honest (what we refer to as genuine) and byzantine identities without defining what they are; here we formalize these notions and provide more sound definitions of harmless and byzantine identifiers. A formal definition of graph conductance $\Phi$ and the proof of Theorem \ref{thm: byz frac} may be found in the supplementary material. 

\begin{remark}
    A potential application of Theorem \ref{thm: byz frac} is a byzantine-resilient union of two communities. Let $A,A'\subseteq V$ denote two communities that have some overlap (non-empty intersection) and wish to unite into $A_2 := A \cup A'$.  Then, if Theorem \ref{thm: byz frac} holds for $(A_1, A_2)$ in case $A_1 := A$ and also in case $A_1 := A'$, this would provide both $A$ and $A'$ the necessary guarantee that the union would not result in an increase of the sybil penetration rate for either community.
\end{remark}

\section{Outlook}

Digital identity  systems face the ``Decentralized Identity Trilemma", of being (\iti) privacy-preserving, (\itii) sybil-resilient and (\itiii) self-sovereign, all at the same time~\cite{DIT}.  It has been claimed that no existing identity system satisfies all three corners of the Trilemma~\cite{DIT}; our approach may be the first to do so.

While this paper provides a formal mathematical framework, we aimed the constructions to be readily amenable to implementation. Realizing the proposed solution entails developing additional components, notably sybil-resilient governance mechanisms, e.g. along the lines of~\cite{SRSC}; a mechanism for encouraging honest behavior and discouraging corrupt behavior, e.g. along the lines of \cite{seuken2014sybil}; and a cryptocurrency to fuel such a mechanism and the system in general.  
%As promising future work, we note the following:
%\begin{enumerate} \item Directed surety graphs, as opposed to the undirected graphs considered here; \item Surety graphs containing surety edges of various types combined;
%\item Agents birth and death, to accommodate a dynamic, real-world setting. \item Relaxations of the surety-based community growth assumptions. \end{enumerate}
Once all components have been designed, we aim to implement, simulate, test, deploy,  and evaluate the proposed framework, hopefully realizing the potential of genuine personal identifiers.

% %
% Note that the difference between a surety graph $G_k$ and its successor $G_{k + 1}$ is either an introduction of a new personal identifier (i.e., vertex) $v$ or a mutual surety edge, or that $G_k$ and $G_{k + 1}$ are equal (this last case happens if the event $c_{k + 1}$ is a surety pledge event which is the first of the two symmetric events).
% %
% While the removal of vertices and/or edges is important, we do not treat these now.
%

%\mypara{Acknowledgements}  We thank the generous support of the Braginsky Center for the Interface between Science and the Humanities and Ouri Poupko for helpful discussions.

%\section*{Acknowledgements}

%We thank the generous support of the Braginsky Center for the Interface between Science and the Humanities.

\newpage
\bibliographystyle{ecai}
\bibliography{bib}

\section{Supplementary Material}

\subsection{Graph notations and terminology}

We provide some more detailed notations regarding graphs and conductance.

Let $G = (V, E)$ be an undirected graph. The \emph{degree} of a vertex $x \in V$ is $\deg(x) := |\{y\in V \> | \> (x,y)\in E\}|$. 
% $G$ is \emph{$d$-regular} if $\deg(x) = d$ holds for each $x \in V$.
%
The \emph{volume} of a given subset $A \subseteq V$ is the sum of degrees of its vertices, $vol(A) := \sum_{x \in A} \deg(x)$. Additionally, we denote the subgraph induced on the set of vertices $A$ as $G|_A$, by $\deg_A(x)$ the degree of vertex $x \in A$ in $G|_A$, and by $vol_A(B):=\sum_{x\in B} \deg_A(x)$ the volume of a set $B \subseteq A$ in $G|_A$. 
Given two disjoint subsets $A,B\subseteq V$, the size of the cut between $A$ and $B$ is denoted by $$e (A,B) = |\{(x,y)\in E ~|~ x\in A,y\in B\}|\ .$$

\mypara{Connectivity measures}
In the following, we define two fundamental notions of graph connectivity that play a substantial role in safe community growth.

\begin{definition}[Combinatorial Conductance]\label{def: conductance}
Let $G = (V, E)$ be a graph. 
The \emph{conductance} of $G$ is defined by:
$$\Phi(G)= \min_{\emptyset\neq A\subset V} \frac{e(A,A^c)}{\min\{vol(A), vol(A^c)\}}\ .$$ ($A^c := V \setminus A$ is the complement of $A$.)
\end{definition}

The following definition is an algebraic measure for connectivity.

\begin{definition}[Algebraic conductance]\label{def: lambda exp}
Let $G$ be a graph, and let $\lambda_n\leq \lambda_{n-1}\leq...\leq \lambda_2\leq \lambda_1$ be the eigenvalues of its random walk matrix. Then, $G$ is a said to be a $\lambda$-expander if its generalized second eigenvalue $\lambda(G):=\max_{i\neq 1}|\lambda_i|$ satisfies $\lambda(G)\leq \lambda$.
\end{definition}

We note that the notions of conductance and algebraic conductance are tightly related via the celebrated Cheeger inequality \cite{cheeger1969lower}. We refer the reader to the text of Hoory et al.~\cite{hoory2006expander} for through exposition and elaborate discussion regarding conductance and graph expansion.

\subsection{Proof of Theorem 2}

Theorem \ref{thm: byz frac} follows by induction from the following Lemma:

\begin{lemma}
Let $G=(V,E)$ be a surety graph with $A\subseteq A'\subseteq V$, and set $\alpha,\beta,\gamma,\delta \in [0,1]$ and $d>0$.\\

Assume:
\begin{enumerate}
    \item $deg(v)\le d$ for all $v\in A'$.
    
    [The graph has a bounded degree].
    
    \item Every $a\in A'$ satisfies $\frac{|\{x\in A'~|~(a,x)\in E\}|}d\geq \alpha$.
    
    [Sufficiently many edges are within members of $A'$]
    
    \item $\frac{|A\cap B|}{|A|}\le\beta$.
    
    [Byzantine penetration to the initial community is bounded]
    
    \item $\frac{e(A'\cap \HC,A'\cap B)}{vol_{A'}(A'\cap \HC)}\leq \gamma$.
    
    [the edges between harmless and byzantine identifiers are relatively scarce]
    
    \item $|A'\setminus A| \leq \delta |A|$, with $\beta + \delta \leq \frac{1}{2}.$
    
    [Community growth is bounded]
    
    \item $\Phi(G|_{A'}) > \frac{\gamma}{\alpha}\cdot \left(\frac{1-\beta}{\beta}\right)$.
    
    [the conductance within $A'$ is sufficiently high]
\end{enumerate}
$\>\>\>$ Then, $\frac{|A'\cap B|}{|A'|}\le\beta$.
\end{lemma}

\begin{proof}
We first note that due to $A\subseteq A'$, and assumptions (3), (5), we have

\begin{align*}
    |A'\cap B| &\le |A\cap B|+|A'\setminus A|\\
    &\le \beta |A| + \delta |A|\\
    &\le \frac{|A|}{2}<\frac{|A'|}{2}.
\end{align*}

As $V=B\uplus H$, it follows that 
\begin{equation} \label{eq: B<H}
    |A'\cap B|<|A'\cap \HC|.   
\end{equation}
We now utilize assumption (1):
\begin{align}\label{eq: vol_A'1}
    vol_{A'}(A'\cap B) &:= \sum_{a\in A'\cap B} |\{x\in A'~|~(a,x)\in E\}| \nonumber \\ 
    &\ge \sum_{a\in A'\cap B} \alpha d = \alpha d|A'\cap B|.
\end{align}

Similarly, we have

\begin{equation} \label{eq: vol_A'2}
    vol_{A'}(A'\cap \HC)\ge\alpha d|A'\cap \HC|.
\end{equation}

Inequalities \ref{eq: vol_A'2} and \ref{eq: B<H} imply that 

$$vol_{A'}(A'\cap \HC)\ge\alpha d|A'\cap B|,$$

and together with Inequality \ref{eq: vol_A'1}, we have:

\begin{equation}\label{eq:min vol}
    \min\{vol(A'\cap \HC), vol(A'\cap B)\} 
    \ge \alpha d|A'\cap B|.
\end{equation}

Now, Inequality \ref{eq:min vol} and assumption (6) imply that:
\begin{align*}
\begin{split}
    \frac{e(A'\cap \HC,A'\cap B)}{\alpha d|A'\cap B|}&\ge\frac{e(A'\cap \HC,A'\cap B)}{\min\{vol(A'\cap \HC), vol(A'\cap B)\}} \\
    &>\frac{\gamma}{\alpha}\cdot \left(\frac{1-\beta}{\beta}\right)\ ,
\end{split}
\end{align*}

or equivalently
\begin{equation}\label{eq: conductance result}
    \frac{e(A'\cap \HC,A'\cap B)}{ d\gamma |A'\cap B|} \ge \frac{1-\beta}{\beta}.
\end{equation}

Assumptions (1) and (4) imply

\begin{equation*}
\frac{e(A'\cap \HC,A'\cap B)}{d|A'\cap \HC|}\leq\frac{e(A'\cap \HC,A'\cap B)}{vol_{A'}(A'\cap \HC)}\leq \gamma\ ,
\end{equation*}

or equivalently

\begin{equation}\label{eq: gamma result}
    |A'\cap \HC| \ge \frac{e(A'\cap \HC,A'\cap B)}{d\gamma}.
\end{equation}

Combining Inequalities \ref{eq: conductance result}, \ref{eq: gamma result} we get:
\begin{align*}
\frac{|A'|}{|A'\cap B|}&=\frac{|A'\cap \HC|+|A'\cap B|}{|A'\cap B|}\\
&\ge\frac{e(A'\cap \HC,A'\cap B)}{d\gamma|A'\cap B|}+1\\
&>\left(\frac{1-\beta}{\beta}\right)+1=\frac{1}{\beta}\ ,
\end{align*}
where the first equality holds as $A=(A\cap \HC)\uplus(A\cap B)$, the second inequality stems from Equation \ref{eq: gamma result} and the third inequality stems from Equation \ref{eq: conductance result}. Flipping the nominator and the denominator then gives $\beta(A'):=\frac{|A'\cap B|}{|A'|}<\beta$.
\end{proof}

\end{document}

\pagestyle{plain}

\title{Supplementary Material for ``Genuine Global Identifiers and their Mutual Sureties''}
\maketitle

\section{Graph notations and terminology}

We provide some more detailed notations regarding graphs and conductance.

Let $G = (V, E)$ be an undirected graph. The \emph{degree} of a vertex $x \in V$ is $\deg(x) := |\{y\in V \> | \> (x,y)\in E\}|$. 
% $G$ is \emph{$d$-regular} if $\deg(x) = d$ holds for each $x \in V$.
%
The \emph{volume} of a given subset $A \subseteq V$ is the sum of degrees of its vertices, $vol(A) := \sum_{x \in A} \deg(x)$. Additionally, we denote the subgraph induced on the set of vertices $A$ as $G|_A$, by $\deg_A(x)$ the degree of vertex $x \in A$ in $G|_A$, and by $vol_A(B):=\sum_{x\in B} \deg_A(x)$ the volume of a set $B \subseteq A$ in $G|_A$. 
Given two disjoint subsets $A,B\subseteq V$, the size of the cut between $A$ and $B$ is denoted by $$e (A,B) = |\{(x,y)\in E ~|~ x\in A,y\in B\}|\ .$$

\mypara{Connectivity measures}
In the following, we define two fundamental notions of graph connectivity that play a substantial role in safe community growth.

\begin{definition}[Combinatorial Conductance]\label{def: conductance}
Let $G = (V, E)$ be a graph. 
The \emph{conductance} of $G$ is defined by:
$$\Phi(G)= \min_{\emptyset\neq A\subset V} \frac{e(A,A^c)}{\min\{vol(A), vol(A^c)\}}\ .$$ ($A^c := V \setminus A$ is the complement of $A$.)
\end{definition}

The following definition is an algebraic measure for connectivity.

\begin{definition}[Algebraic conductance]\label{def: lambda exp}
Let $G$ be a graph, and let $\lambda_n\leq \lambda_{n-1}\leq...\leq \lambda_2\leq \lambda_1$ be the eigenvalues of its random walk matrix. Then, $G$ is a said to be a $\lambda$-expander if its generalized second eigenvalue $\lambda(G):=\max_{i\neq 1}|\lambda_i|$ satisfies $\lambda(G)\leq \lambda$.
\end{definition}

We note that the notions of conductance and algebraic conductance are tightly related via the celebrated Cheeger inequality \cite{cheeger1969lower}. We refer the reader to the text of Hoory et al.~\shortcite{hoory2006expander} for through exposition and elaborate discussion regarding conductance and graph expansion.

\section{Proof of Theorem 2}

Theorem \ref{thm: byz frac} follows by induction from the following Lemma:

\begin{lemma}
Let $G=(V,E)$ be a surety graph with $A\subseteq A'\subseteq V$, and set $\alpha,\beta,\gamma,\delta \in [0,1]$ and $d>0$.\\

Assume:
\begin{enumerate}
    \item $deg(v)\le d$ for all $v\in A'$.
    
    [The graph has a bounded degree].
    
    \item Every $a\in A'$ satisfies $\frac{|\{x\in A'~|~(a,x)\in E\}|}d\geq \alpha$.
    
    [Sufficiently many edges are within members of $A'$].
    
    \item $\frac{|A\cap B|}{|A|}\le\beta$.
    
    [Byzantine penetration to the initial community is bounded.]
    
    \item $\frac{e(A'\cap \HC,A'\cap B)}{vol_{A'}(A'\cap \HC)}\leq \gamma$.
    
    [the edges between harmless and byzantine identifiers are relatively scarce]
    
    \item $|A'\setminus A| \leq \delta |A|$, with $\beta + \delta \leq \frac{1}{2}.$
    
    [Community growth is bounded.]
    
    \item $\Phi(G|_{A'}) > \frac{\gamma}{\alpha}\cdot \left(\frac{1-\beta}{\beta}\right)$.
    
    [the conductance within $A'$ is sufficiently high]
\end{enumerate}
$\>\>\>$ Then, $\frac{|A'\cap B|}{|A'|}\le\beta$.
\end{lemma}

\begin{proof}
We first note that due to $A\subseteq A'$, and assumptions (3), (5), we have

\begin{align*}
    |A'\cap B| &\le |A\cap B|+|A'\setminus A|\\
    &\le \beta |A| + \delta |A|\\
    &\le \frac{|A|}{2}<\frac{|A'|}{2}.
\end{align*}

As $V=B\uplus H$, it follows that 
\begin{equation} \label{eq: B<H}
    |A'\cap B|<|A'\cap \HC|.   
\end{equation}
We now utilize assumption (1):
\begin{align}\label{eq: vol_A'1}
    vol_{A'}(A'\cap B) &:= \sum_{a\in A'\cap B} |\{x\in A'~|~(a,x)\in E\}| \nonumber \\ 
    &\ge \sum_{a\in A'\cap B} \alpha d = \alpha d|A'\cap B|.
\end{align}

Similarly, we have

\begin{equation} \label{eq: vol_A'2}
    vol_{A'}(A'\cap \HC)\ge\alpha d|A'\cap \HC|.
\end{equation}

Inequalities \ref{eq: vol_A'2} and \ref{eq: B<H} imply that 

$$vol_{A'}(A'\cap \HC)\ge\alpha d|A'\cap B|,$$

and together with Inequality \ref{eq: vol_A'1}, we have:

\begin{equation}\label{eq:min vol}
    \min\{vol(A'\cap \HC), vol(A'\cap B)\} 
    \ge \alpha d|A'\cap B|.
\end{equation}

Now, Inequality \ref{eq:min vol} and assumption (6) imply that:
\begin{align*}
\begin{split}
    \frac{e(A'\cap \HC,A'\cap B)}{\alpha d|A'\cap B|}&\ge\frac{e(A'\cap \HC,A'\cap B)}{\min\{vol(A'\cap \HC), vol(A'\cap B)\}} \\
    &>\frac{\gamma}{\alpha}\cdot \left(\frac{1-\beta}{\beta}\right)\ ,
\end{split}
\end{align*}

or equivalently
\begin{equation}\label{eq: conductance result}
    \frac{e(A'\cap \HC,A'\cap B)}{ d\gamma |A'\cap B|} \ge \frac{1-\beta}{\beta}.
\end{equation}

Assumptions (1) and (4) imply

\begin{equation*}
\frac{e(A'\cap \HC,A'\cap B)}{d|A'\cap \HC|}\leq\frac{e(A'\cap \HC,A'\cap B)}{vol_{A'}(A'\cap \HC)}\leq \gamma\ ,
\end{equation*}

or equivalently

\begin{equation}\label{eq: gamma result}
    |A'\cap \HC| \ge \frac{e(A'\cap \HC,A'\cap B)}{d\gamma}.
\end{equation}

Combining Inequalities \ref{eq: conductance result}, \ref{eq: gamma result} we get:
\begin{align*}
\frac{|A'|}{|A'\cap B|}&=\frac{|A'\cap \HC|+|A'\cap B|}{|A'\cap B|}\\
&\ge\frac{e(A'\cap \HC,A'\cap B)}{d\gamma|A'\cap B|}+1\\
&>\left(\frac{1-\beta}{\beta}\right)+1=\frac{1}{\beta}\ ,
\end{align*}
where the first equality holds as $A=(A\cap \HC)\uplus(A\cap B)$, the second inequality stems from Equation \ref{eq: gamma result} and the third inequality stems from Equation \ref{eq: conductance result}. Flipping the nominator and the denominator then gives $\beta(A'):=\frac{|A'\cap B|}{|A'|}<\beta$.
\end{proof}

% \begin{remark}
% To internalize the above definition, consider the options available to a community that wishes to achieve resilience to a given Byzantine penetration~$\beta$:
% (\iti) to increase solidarity, $\alpha$, up to 1; (\itii) to increase conductance, $\Phi$, up to $\nicefrac{1}{2}$; (\itiii) or to curb the attack, $\gamma$, down to 0. However: (\iti) increasing solidarity may prevent the community from admitting additional members, defeating the purpose of community growth; (\itii) increasing conductance may require community members who do not know each other to trust each other, defeating the notion of trust;
% and
% (\itiii) curbing the attack will require the community to expose sybils and their corrupt perpetrators, which is an effort.
% %
% In summary, achieving $(\alpha,\beta,\gamma)$-resilience for a given $\beta$ is a challenge, and addressing it will require a community to craft a balance between its desire to grow, the measures it is willing to undertake to support such growth, %e.g. expenditures on incentives and assessing penalties,
% and the risks it is willing to undertake in order to grow.  %Clearly, in an emergency, such as the detection of a sybil attack, growth should be suspended until enough sybils and their corrupt perpetrators are detected, the sybils eradicated and their perpetrators neutralized.

% \end{remark}

%

\bibliographystyle{aaai}
\bibliography{bib}